# Relationship between acoustic body waves and *in situ* stresses around a borehole

André Rousseau

*CNRS-UMS 2567 (OASU)*
*Université Bordeaux 1 - Groupe d'Etude des Ondes en Géosciences*
*351, cours de la Libération,F-33405 Talence cédex*
a.rousseau@geog.u-bordeaux1.fr

**Summary**

This paper shows that there are three types of shape of acoustic body waves propagating *inside competent and homogeneous formations* penetrated by a borehole : simple, double, and resonant. This fact is connected to the modifications undergone by the area located around a well after drilling in relation to *in situ* state of stress. There are overstressed and understressed vertical cylindrical zones with "bubble-shaped" cross-sections, in which acoustic body waves are supposed to reflect. The horizontal size and the shape of the stress modified areas depend on the values of $(K_1+K_2)/2$ - with $K_1=\sigma_1$(or $\sigma_H$)/$\sigma_v$ (the overburden pressure) and $K_2=\sigma_2$(or $\sigma_h$)/$\sigma_v$ -, that is to say on the *in situ* horizontal stress and the anisotropy of this stress. The calculation of the velocities of the P and S double waves allows us to evaluate the radial thicknesses of these stress modified areas. As their values are different according to whether they result from P or S waves, we propose an explanation based on their wavelengths. The comparisons with other works on the *in situ* stress of the studied boreholes allow us to propose a method for evaluating the horizontal stress and its anisotropy, thanks to the estimation of the value $(K_1+K_2)/2$.

**Keywords**
Borehole, acoustic body waves, *in situ* stress, anisotropy.

**Introduction**

The drilling of a borehole involves stress modifications around the borehole possibly leading to various rupture modes (see Maury, 1987, 1992) now observed with borehole imagery tools (see Maury et al., 1999) ; those modifications strongly depend on the anisotropy of *in situ* horizontal state of stress, and also on depth. The area surrounding the well undergoes stress modifications resulting in overtressed and understressed zones according to the azimuth of the considered point on the periphery of the hole related to the main horizontal stress components. When this stress redistribution exceeds the failure criterion of the rock, one of the well known consequences is induced wall fractures, such as breakouts. Before the onset of rupture, rock behaves elastically, and at the borehole wall as well as inside the formation, a sharp modification of the stress occurs. This stress variation alters the body wave velocities, and may involve reflection and refraction phenomena, if the wavelengths are compatible, which is the case of acoustic waves (about 20 kHz).

We have pointed out that the acoustic waveforms obtained in deep boreholes crossing *competent and homogeneous formations* scarcely display simply shaped body waves when they are rapid : it is particularly true in the case of P waves, where we can observe many times the repetition of particular major "crests". However there is generally no surrounding reflection plane, which might explain such wave characteristics (Rousseau, 2000, 2001). On the other hand, still in competent formations (for example, crystalline rocks), body waves may exhibit a kind of resonance. We assume in this paper that those observations can be explained by a common origin linked to *in situ* state of stress.

The acoustic data of this study come, on the one hand, from three deep scientific cored boreholes : the KTB Pilot Hole in Bavaria (Germany) drilled inside crystallophyllian rocks, Balazuc1 (South of France) inside compact sandstones and carbonated rocks, and GPK1 in Soultz-sous-forêt (Alsace, France) inside granites ; and on the other hand, from the ANDRA boreholes in the Vienne region (France), drilled and cored inside granites at shallower depths.

**The different shapes of the body waves and stress changing around a borehole**

The acoustic body waves (source of 20 kHz) propagating rapidly within homogeneous and competent formations exhibit different shapes, which can be classified into three cases. Fig. 1 shows an example of each of them :

1) the simplest and basic form,
2) the "double" P and S waves (as an echo of each first wave arrival),
3) the "resonance" form.

The occurrence of "double" waves inside an homogeneous formation has been supposed to be caused by the *stress modifications* involved around a hole after drilling in competent formations, without thermal problems (see Maury & Guénot, 1988 ; Maury & Idelovici, 1995), and with Biot's coefficient of zero.

A competent formation reacts to a drilling in relation to *in situ* state of stress. The modifications undergone by the area surrounding a well are complex (see the studies by Maury et al, 1999, in relationship with rupture modes observed by logging tools). The velocity of acoustic body waves propagating through a borehole will be modified, if the areas undergoing a large stress modification are thicker than the wavelengths.

Let a vertical hole be drilled in a vertical and horizontal principal stress field, a perfectly compact rock (effective stress equal to total stress) with linear elastic isotropic behaviour, a stabilised new state of stress in an anisotropic medium ($\sigma_1$=maximum horizontal stress ; $\sigma_2$=minimum horizontal stress), and negligible thermal effects ; the radial $\sigma_r$ and tangential $\sigma_\theta$ stresses are then :

$$\sigma_r = \frac{\sigma_1 + \sigma_2}{2}\left(1 - \frac{a^2}{r^2}\right) + \frac{\sigma_1 - \sigma_2}{2}\left(1 - \frac{4a^2}{r^2} + \frac{3a^4}{r^4}\right)\cos 2\theta + P_m\frac{a^2}{r^2}$$

$$\sigma_\theta = \frac{\sigma_1 + \sigma_2}{2}\left(1 + \frac{a^2}{r^2}\right) - \frac{\sigma_1 - \sigma_2}{2}\left(1 + \frac{3a^4}{r^4}\right)\cos 2\theta - P_m\frac{a^2}{r^2}$$

with $a$ the hole radius and $r$ the distance to the hole centre in the horizontal plane, and $\theta$ the azimuth in relation to the direction of $\sigma_1$ : $P_m$ is the mud or water pressure of the hole.

The vertical stress $\sigma_v$ deduced by the condition of plane deformation is the lithostatic pressure and $\Delta\sigma_v$ the variation of the vertical stress in relation to $\theta$ and $\nu$ the Poisson ratio :

$$\Delta\sigma_v = -\nu\frac{4a^2}{r^2}\frac{\sigma_1 - \sigma_2}{2}\cos 2\theta$$

The mean values of the steady stress modifications involved by the drilling are expressed by $\Delta\sigma = (\Delta\sigma_r + \Delta\sigma_\theta + \Delta\sigma_v)/3$, where $\Delta\sigma_r$ and $\Delta\sigma_\theta$ are respectively the differences before and after drilling of $\sigma_r$ and $\sigma_\theta$ ($a = r = 0$ and $\theta = 0$ in the first case). Some examples of the distribution of their isovalue curves in relation to hole radius and over a given distance around and from the hole are visible in Figure 6 in function of the parameters $K_1 = Q_1/\sigma_v$ and $K_2 = Q_2/\sigma_v$ with : $Q_1 = \sigma_1$ and $Q_2 = \sigma_2$, the horizontal stress anisotropy being equal to $(Q_1 - Q_2)/[(Q_1 + Q_2)/2]$. These parameters are conspicuous because we will admit that :

1) $(K_1+K_2)/2 = 0.7$ tends to represent a medium of weak or tensile stresses,
2) $(K_1+K_2)/2 = 1$ tends to represent a medium of constraint stresses,
3) $(K_1+K_2)/2 = 1.3$ tends to represent a medium of shear stresses.

As the wavelengths of acoustic body waves range from 0.25 m for P waves to 0.17 m for S waves, the size of the stress modified areas is compatible with the reverberation of acoustic waves within those areas.

If there is a reflection inside a stress modified area, its size can be evaluated from the calculation of the different velocities of all the acoustic body waves encounted in full waveforms. This is made from the move-out displayed by the tracks of the aligned receivers of a monopole probe.

**"Double" P and S waves ; the "crest" velocities**

Figure 2 shows some waveforms drawn from the studied boreholes and selected in areas without fracture plane nor any other reflection plane. We can see two separate P waves and two separate S waves, each of them being characterised by a major "crest" whose the mean velocity of each has been calculated from their move-out.

**Evaluation of the "radial thicknesses" of the stress modified areas**

Assuming the *hypothesis* of multiple reflections occurring inside the over and under stressed areas located around a well, the second "crest" of the double "crests" represents the first multiple of the corresponding body wave. The thickness e of the modified area can easily be calculated as

$$e = \tfrac{1}{2} ((t_{c2} - t_{c1}) * V_{c2})$$

where $V_{c2}$ is the velocity of the second "crest" and $(t_{c2} - t_{c1})$ the time measured between both "crests" of a pair of "crests".

For the studied boreholes, we have calculated the distribution of the radial thickness values of the stress modified areas for P and S waves, as well as the one of the resonance occurrence – in the wells where resonance appears -, as a function of depth over 50 m by the mean of the moving average.

Figures 3 and 4 show that there are two main kinds of distribution of the radial thickness values in relation to depth : these values are either *scattered* (case of the Balazuc1 borehole, and the KTB Pilot Hole above 3000 m.), or *steady* (case of the GPK1 borehole (Soultz) and the KTB Pilot Hole below 3000 m.). The distributions in percentage of the calculated radial thickness values over a "moving" vertical space of 50 m are represented in Fig. 4 for both those calculated from P waves and those from S waves. In order to increase the reliability of these calculations, the radial thickness values which give Fig. 3 and Fig. 4 were taken into account *only if there were values available at the same depth from P and S waves*.

Concerning the P wave results, *the zones of scattered values* display radial thickness values which range *from 0.2 m to 0.7 m* (see Fig. 4), *the zones of steady values* display radial thickness values of *0.3 and 0.4 m.* As for the S wave results, *the zones of scattered values* display radial thickness values which range *from 0.4 to 0.7 m with a zone of varying values from 0.3 to 0.5* and *the zones of steady values* of *0.5 m and 0.2 m.* So, we observe that *(i)* the P and S waves provide different results, and *(ii)* the zones of scattered values are above the zones of steady values.

In the case of the ANDRA boreholes (Figure 5), which are not as deep, when the body waves are not "resonant", the "double" crests provide, from the P waves, *steady values* of *0.3 m* of the radial thickness below 150-250 m of depth, and *scattered values* ranging *from 0.1 to 0.4 m* above that depth. As for the S waves, they provide only *scattered values* ranging *from 0.1 to 0.5 m,* which tend to ranging *from 0.1 to 0.3 m* above 200 m of depth.

**Discussion and comparisons with the *in situ* stresses studied by other methods**

As the frequency of the acoustic waves usually used is 20 kHz, with wave lengths ranging consequently from about 0.17 m for S waves to 0.3 m for P waves, the stress modified areas are not wide enough to cause the vertical propagation of refracted waves inside a continuous medium.

Two problems arise from the previous results :
1)  Why are there different values according as they are provided from P or S waves ?
2)  Why are there "steady" and "scattered" values ?

It has been assumed that the wavelength does determine the location of the "reverberation" inside the stress modified areas in function of $|\Delta\sigma|$.

According to the calculations of the *in situ* stress by Zang et al. (1990) in the KTB Pilot Hole, the vertical stress $\sigma_v$ would be approximately equal to the maximum horizontal stress $\sigma_H$ at 3000 m of depth, that is to say about 81 Pa, although Roeckel & Natau (1993) calculated a higher magnitude of $\sigma_h$ - more than 102 MPa - at 3011 m of depth, derived from the data of hydraulic fracturing. This corresponds to the sharp limit at this depth between the zones where $\sigma_H > \sigma_v > \sigma_h$ provide the scattered values of the modified area thicknesses, and the zones where $\sigma_v > \sigma_H > \sigma_h$ provide the steady values.

In the GPK1 borehole, the *in situ* horizontal stresses were indirectly approached from hydraulic-fracturing stress measurements by - among several authors - Cornet et al. (1997), and from comparisons with existing stress data of the surrounding tectonic units by Klee & Rummel (1999). The first authors estimate at 2900 m of depth the equality between the vertical stress and the maximum horizontal stress, and the second authors at 3000 m. As a first approach, this limit seems to be confirmed in Fig. 4, where a change in the distribution of the radial thickness values is visible at 2840 m of depth of the GKP1 borehole. Cornet et al. (1997) estimate the minimum horizontal stress between 40 and 45 MPa at 2900 m, and Klee & Rummel (1999) give the values of 23.5 and 17.7 MPa for respectively the maximum and minimum horizontal stresses at 1458 m. The zone located between 2000 and 3500 m of depth may be considered as a zone of steady radial thickness values, the distribution of which being unimodal but varying in relation to depth because of the number of fractures (reservoir zone) (Genter et al., 1997).

The results obtained from S waves in the zones of steady radial thickness values range from mostly 0.5 m in the GPK1 borehole to 0.2 m in the KTB Pilot Hole. This difference might be attributed to the existence of *breakouts* in the KTB Pilot Hole, particularly below 3000 m depth (Kück, 1993). In a first approach, the presence of microcracks might cause the scattering of the radial thickness values of the stress modified areas.

Figure 6 shows models of horizontal stresses integrating some of our calculations of radial thicknesses in the steady values zones and the corresponding values provided by literature. So, in Figure 6a corresponding to the GPK1 borehole, at 2900 m of depth, where $K_1$ is supposed to be equal to 1 and therefore $Q_1 = 78.3$ Mpa, and $Q_2 = 40$ Mpa, the estimations of Cornet et al. (1997) drawn from hydraulic fracturing measurements provide an horizontal stress anisotropy superior to 50 %, and Klee & Rummel (1999) propose an anisotropy of 66 % at 3000 m of depth. In our model, the anisotropy is 65 %. The radial thickness of 0.3 m where P wavelength (0.3 m) can allow a reflection roughly corresponds to $|\Delta\sigma| = 1$ Mpa, whereas the radial thickness of 0.5 m where S wavelength (0.175 m) can allow a reflection roughly corresponds to $|\Delta\sigma| = 0.5$ Mpa (border of the blue surface).

In the case of the KTB Pilot Hole, Fig. 6b shows the theoretical model at 3000 m of depth for $K_1 = 1$, therefore $Q_1 = 81$ Mpa (see above). We must assume $Q_2 = 35$ Mpa in order to make P wavelength corresponding to the radial thickness of 0.3 m and the 1 Mpa curve. Bücher et al. (1990) estimate a general anisotropy of 30 % for practically all the physical parameters (but without stress), even up to 40 % for the radial P wave velocities in cores.

According to the calculations of Zang et al. (1990) from cores, the horizontal stress anisotropy is about 30 %, up to 60 %. For Roeckel and Natau (1993), the stress values inferred from hydraulic fracture radial measurements supply an horizontal stress anisotropy superior to 70 % at 3011 m of depth. This study concludes to an anisotropy of 79 %. The radial thickness value of 0.2 m provided by S waves is less than the theoretical one (about 0.5 m), which can be explained by the presence of the vertical fractures (the breakouts) which divide radially in two parts the overstressed areas (*i-e* the fastest ones), while S waves propagate by radial shear.

Fig. 6c shows the model proposed for the ANDRA boreholes : it "needs" a high anisotropy because of shallow data (500 m of depth). It is perhaps a cause of resonance.

**Conclusion**

The propagation of acoustic body waves through a vertical hole drilled inside competent and fast formations - sedimentary, granitic or metamorphic - may appear under three shapes : basic (rare), split, or resonant. The goal was to check whether there is a correlation between this phenomenon and the stress modifications occurring around a hole after drilling, with an overstressed modified area under the azimuth of the minimum horizontal stress and an understressed modified area under the azimuth of the maximum horizontal stress. The second wave of a pair is then interpreted as a multiple inside those stress modified areas.

This correlation appears relevant from the calculation of the velocities of P and S acoustic waves and of their "doubles" interpreted as their multiples inside the stress modified areas around a hole. This calculation was applied to the log data of several wells and the results prove to be consistent. The value of $(K_1+K_2)/2$ and *in situ* horizontal stress anisotropy governs the radial thickness values of the stress modified areas, but there may be two domains : above a certain depth limit the radial thickness values for both P and S waves appear scattered between 0.1 and 0.7 m, whereas they are steady below that limit. The comparison with the other works about the *in situ* stress of the studied wells tends to the conclusion that the limit dividing the two zones may then represent the depth where there would be equality between maximum horizontal stress and overburden pressure (easy to calculate).

The radial thickness values are not identical, at a given depth, according as they are calculated from P or S waves. It has been deduced from our data that it is the consequence of the differences in the wavelength of P and S waves, their reflection inside the stress modified areas being at different locations.

It is then possible to use those phenomena in order to evaluate *in situ* horizontal stress from the determination of $(K_1+K_2)/2$ in the domain of the steady radial thickness values of the stress modified areas.

**ACKNOWLEDGEMENT**

The author wants to express his gratitude to Kurt Bram (G.G.A., Hanover), Albert Genter (B.R.G.M.), and to the A.N.D.R.A. for their kind and efficient help in collecting the data and for the fruitful discussions with them. He is particularly grateful to Vincent Maury, expert in well mechanics, whose collaboration has permitted this work.

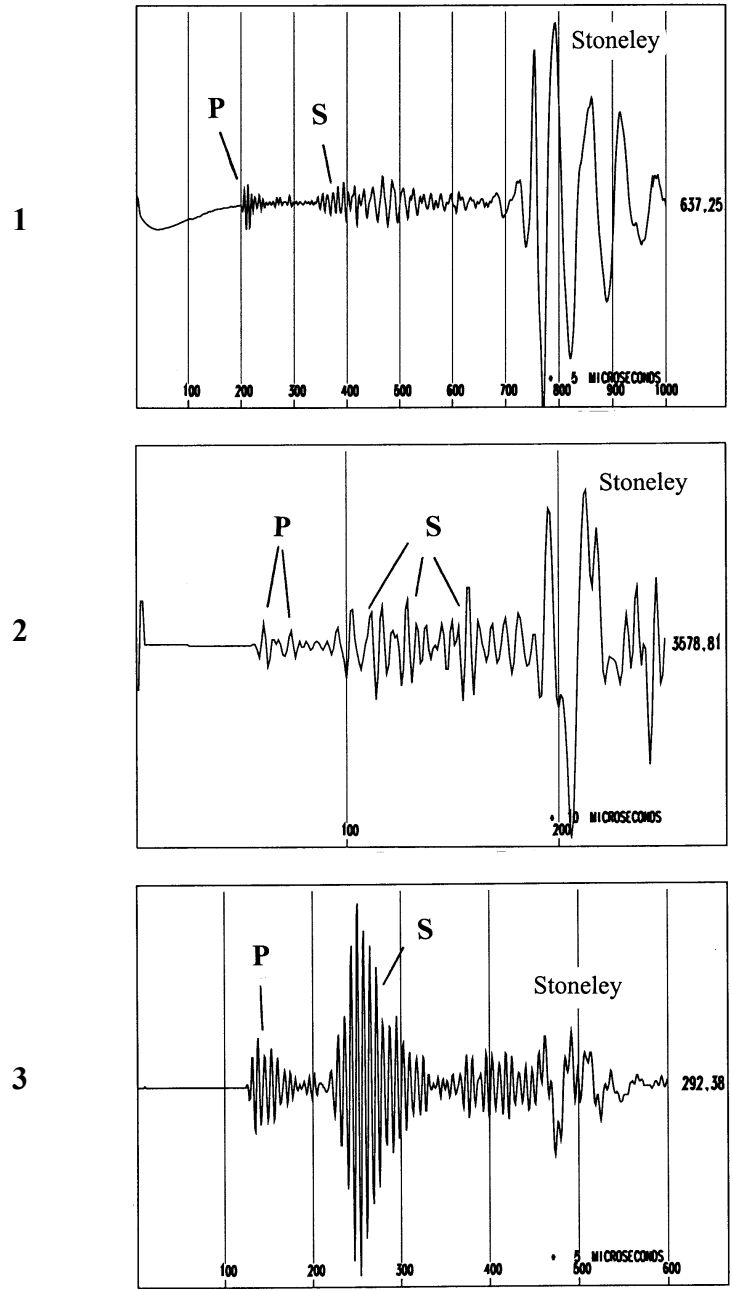

**Figure 1 :** The different shapes of body waves of acoustic full waveforms :

    1) the simplest and basic shape (from the Auriat borehole in the Massif Central, France),

    2) the split shape (multiples) (from the KTB Pilot Hole in Bavaria, Germany),

    3) the resonant form (from the ANDRA boreholes in the Vienne region, France).

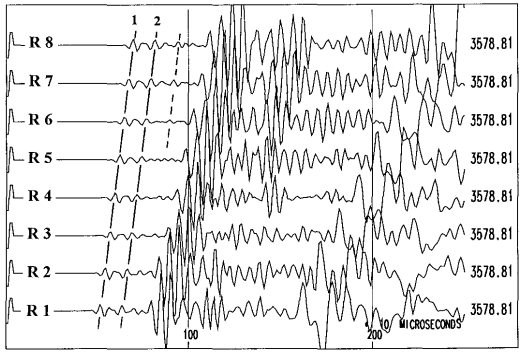

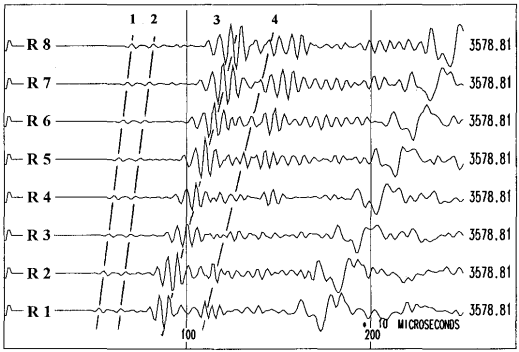

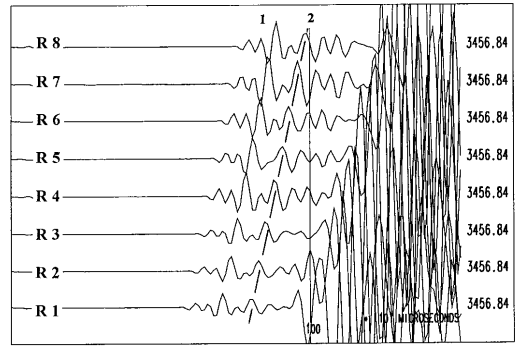

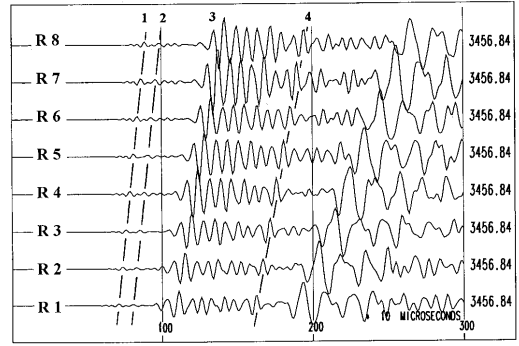

**a**

**b**

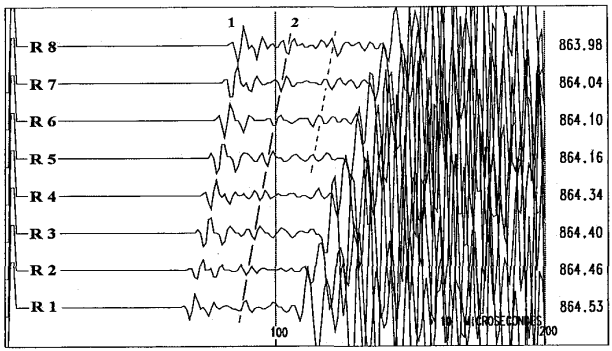

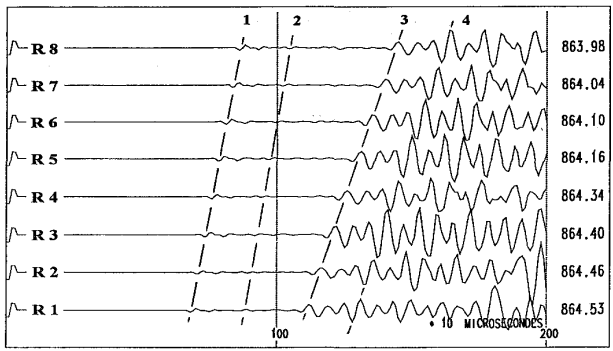

**c**

**Figure 2-1 :** P (numbers 1 and 2) and S (numbers 3 and 4) waves of the full waveforms recorded by the eight receivers ($R_1$ to $R_8$) of a sonic probe :

- a) *in the KTB Pilot Hole (Bavaria, Germany),*
- b) *in the GPK1 borehole (Soultz-sous-Forest, Alsace, France),*
- c) *in the Balazuc1 borehole (South of France).*

Depths are indicated in meters on the right ; the horizontal axis indicates the number of sample times (every 10 µs.) For each example, there are two displays of the same data, but with a different amplitude gain because of the large difference of energy between the P and S waves.

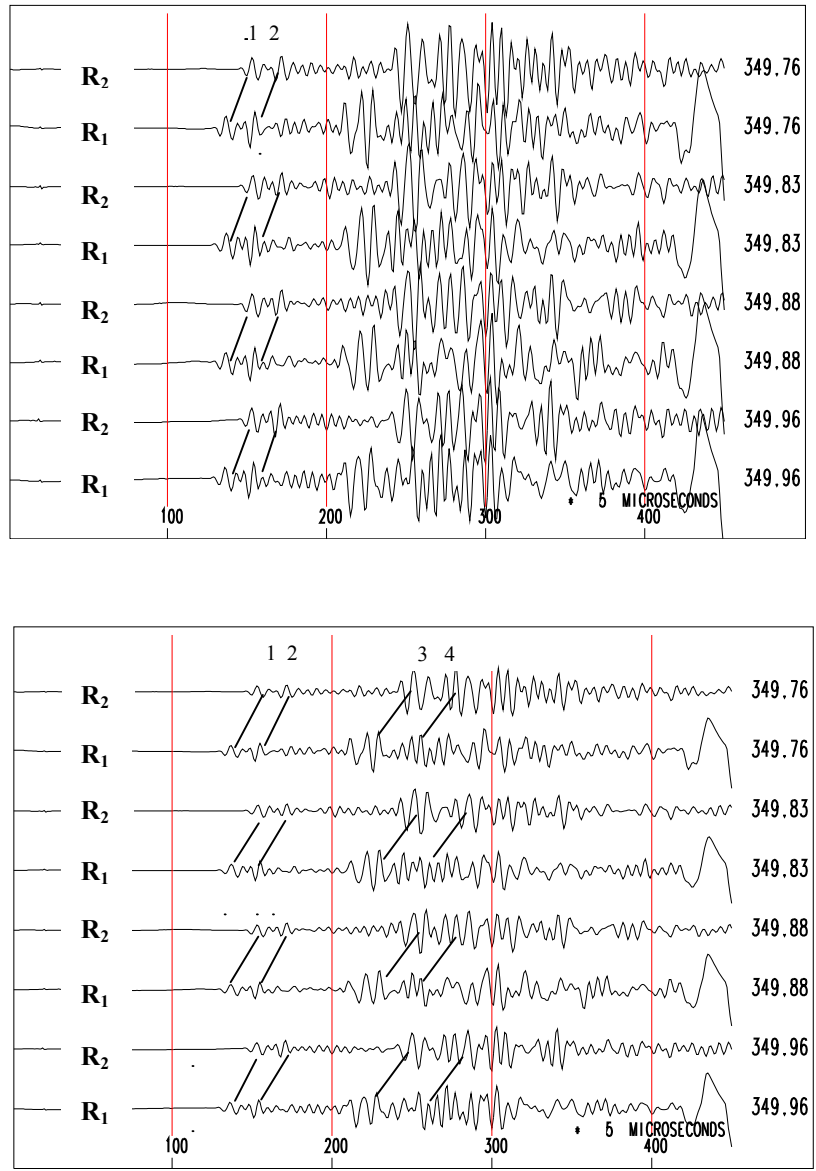

**Figure 2-2 :** P (numbers 1 and 2) and S (numbers 3 and 4) waves of the full waveforms recorded by the two receivers ($R_1$ and $R_2$) of a sonic probe in the ANDRA CHA106 borehole (West of France).

There are four successive depths indicated in meters on the right ; the horizontal axis indicates the number of sample times (every 5 μs.) For each example, there are two displays of the same data, but with a different amplitude gain.

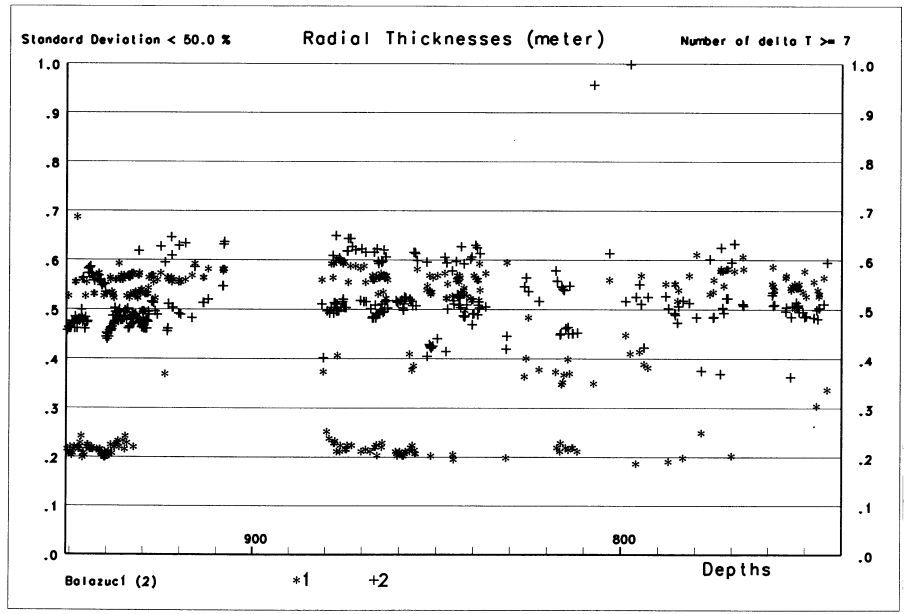

**a**

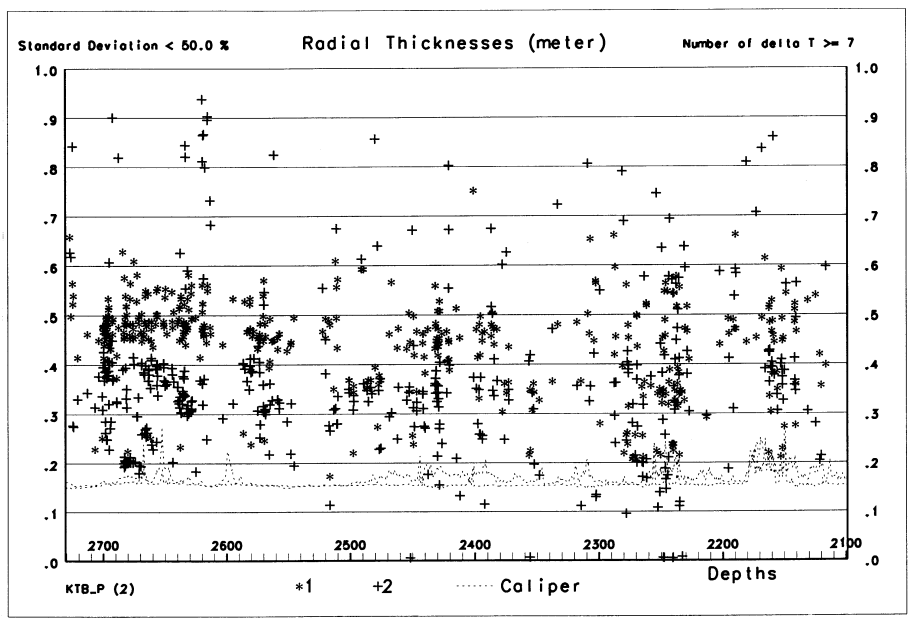

**b**

**Figure 3 :** Examples of "radial thickness" values in relation to depth, drawn from four different areas ; the stars (no 1) represent the radial thickness values calculated from P waves, and the crosses (no 2) those calculated from S waves ; the dashed lines represent the calipers.

  *1) Zones of scattered values :*
   **a)** Balazuc1 borehole,
   **b)** KTB Pilot borehole above 3000 meters,

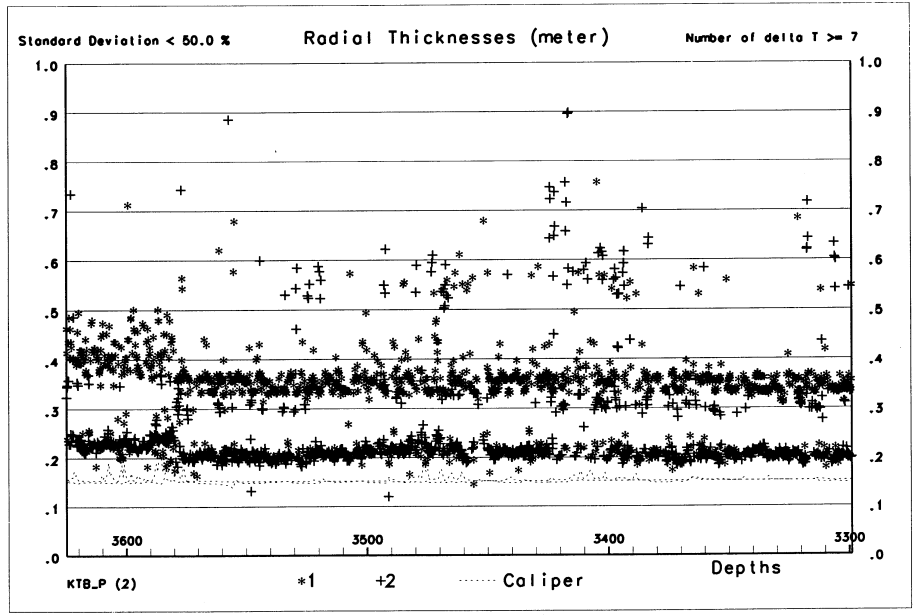

**a**

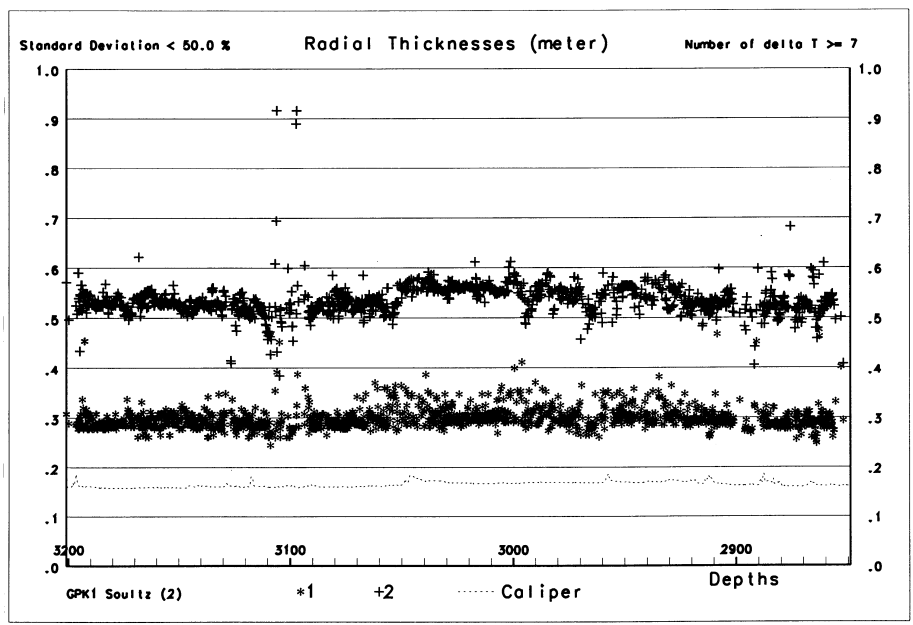

**b**

**Figure 3 :** Examples of "radial thickness" values in relation to depth, drawn from four different areas ; the stars (nb 1) represent the radial thickness values calculated from P waves, and the crosses (nb 2) those calculated from S waves ; the dashed lines represent the calipers.

    *2)    Zones of steady values* :
        **a)** KTB Pilot borehole below 3000 meters,
        **b)** GPK1 borehole (Soultz

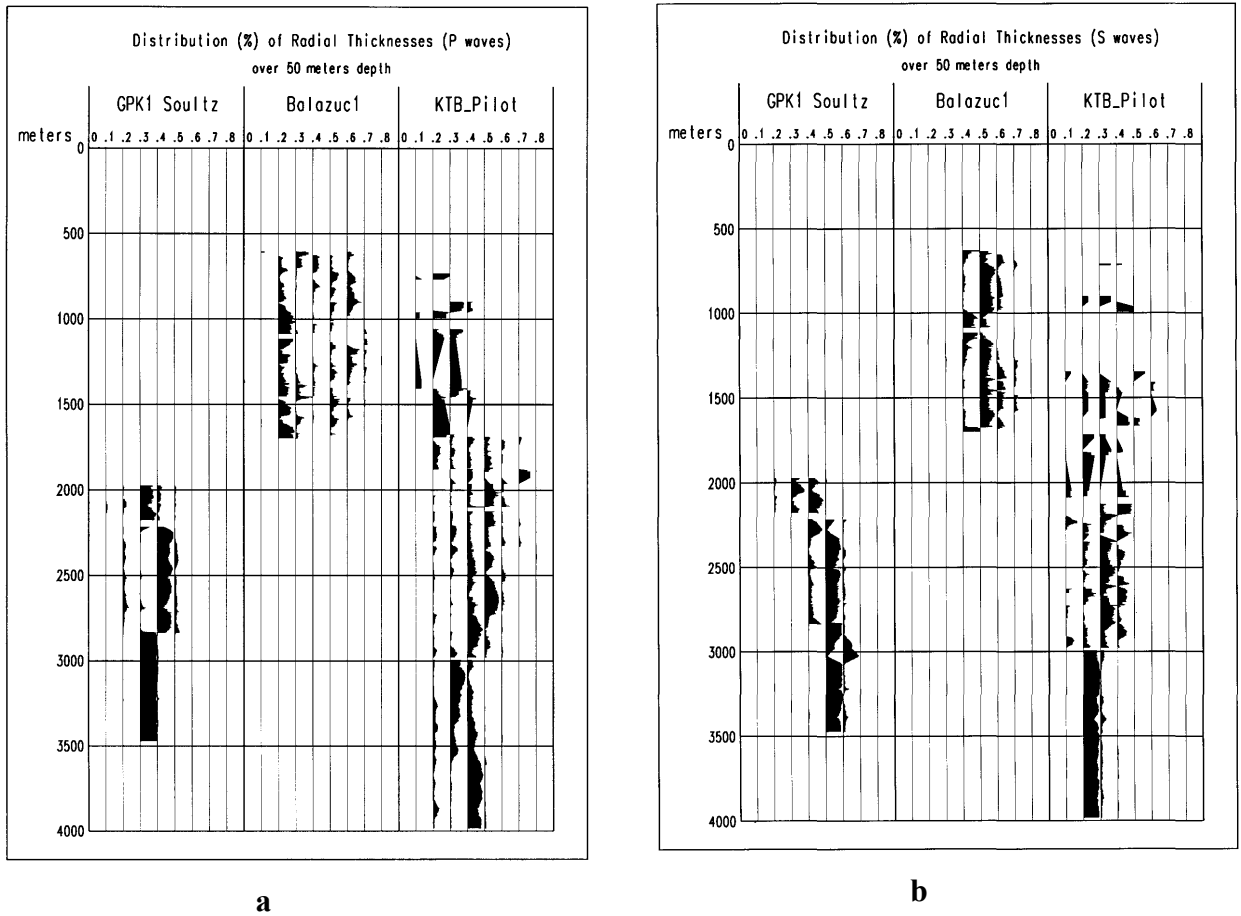

a                                             b

**Figure 4 :** Distribution in percent of the radial thickness values calculated
*a) from P waves,*                                     *b) from S waves,*
in three study deep boreholes over a moving vertical space of 50 m.

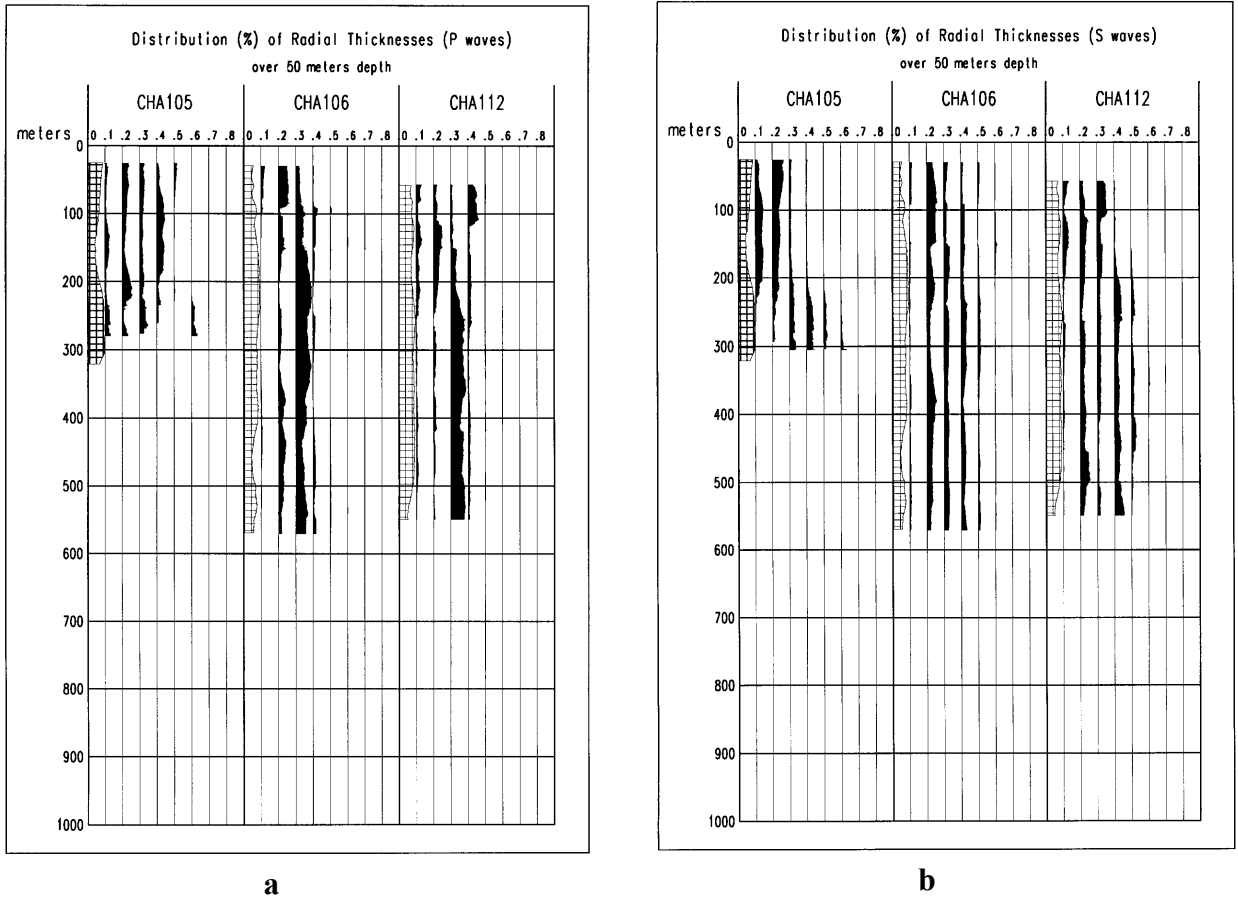

**Figure 5 :** Distribution in percent of the resonance and of the radial thickness values calculated

*a) from P waves,*                    *b) from S waves,*

in three ANDRA boreholes over a moving vertical space of 50 m.

The percentage of *resonance* is indicated by the areas marked out in squares (the full interval between two vertical lines represents 100 %)

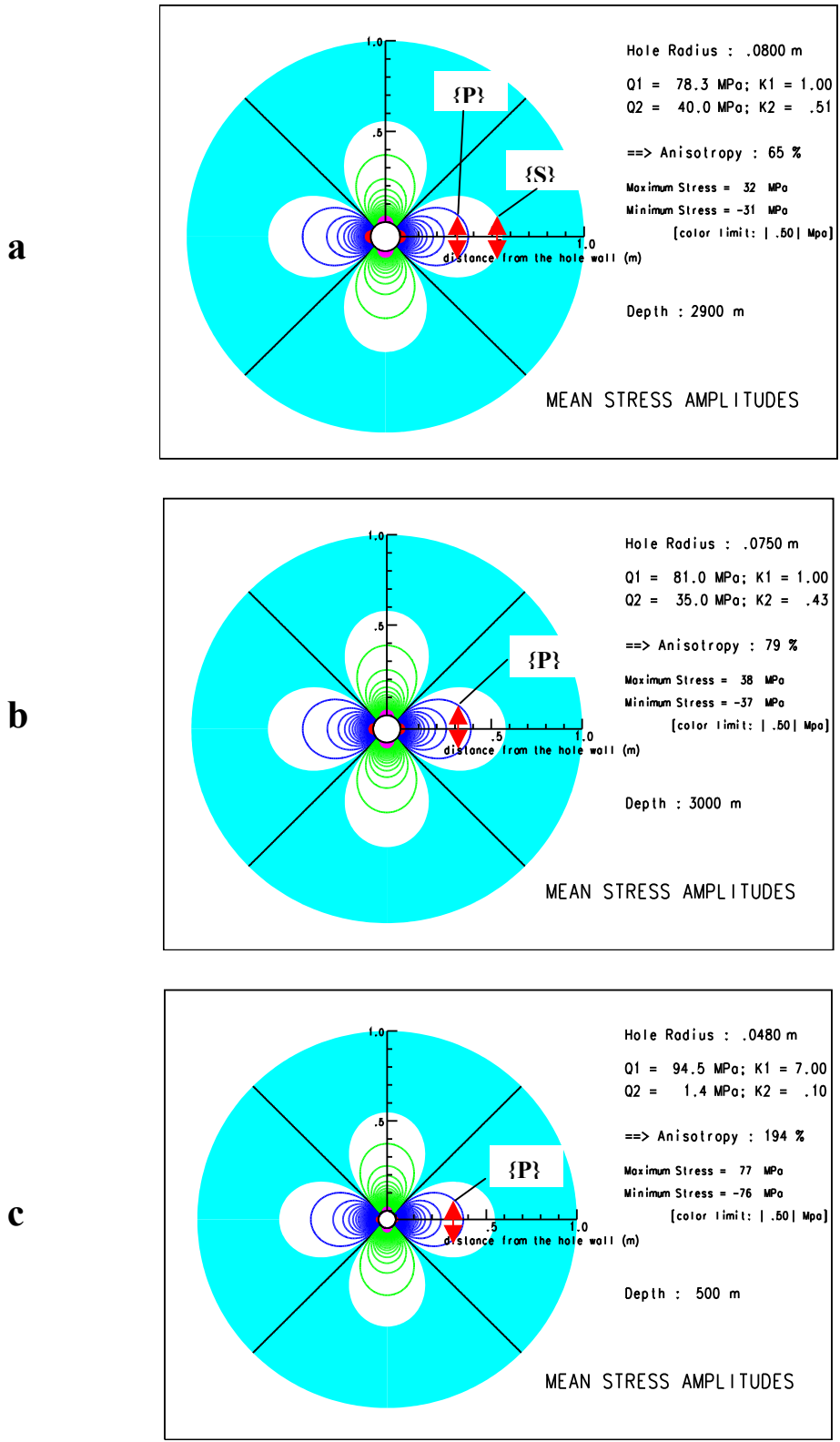

**Figure 6 :** possible models corresponding to : **a)** *the GPK1 borehole (Soultz),* **b)** *the KTB Pilot Hole,* **c)** *the ANDRA boreholes in the Vienne region.*

The blue and green curves indicate respectively the positive and negative isovalues in MPa of $\Delta\sigma$, the black curve the value equal to zero. For a better visualisation, the stress value curves superior to the half of the maximum stress are red, and the curves inferior to the half of the minimum stress are purple. The light blue surface represents the area where $|\Delta\sigma|$ tends to zero from the conventional absolute value 0.5 MPa.